\documentclass{emulateapj}
\usepackage{verbatim,graphicx,natbib}
\usepackage{amsmath}

\newcommand{\exocart}{\texttt{exocartographer}}

\shortauthors{Farr et al.}
\shorttitle{exocartographer}
\begin{document}

\title{\texttt{exocartographer}: A Bayesian Framework for Mapping Exoplanets in Reflected Light}
\author{Ben Farr\altaffilmark{1,2}, Will M.\ Farr\altaffilmark{3}, Nicolas B.\ Cowan\altaffilmark{4}, Hal M. Haggard\altaffilmark{5,6}, Tyler Robinson\altaffilmark{7}}

\altaffiltext{1}{Department of Physics, University of Oregon, Eugene, OR 97403, USA}
\altaffiltext{2}{Enrico Fermi Institute, University of Chicago, Chicago, IL 60637, USA}
\altaffiltext{3}{Birmingham Institute for Gravitational Wave Astronomy, School of Physics and Astronomy, University of Birmingham, Birmingham B15 2TT, United Kingdom}
\altaffiltext{4}{Department of Physics and Department of Earth \& Planetary Sciences, McGill University, 3550 rue University, Montr\'eal, QC H3A 2A7, Canada}
\altaffiltext{5}{Physics Program, Bard College, 30 Campus Road, Annondale-On-Hudson, NY 12504, USA}
\altaffiltext{6}{Perimeter Institute for Theoretical Physics, 31 Caroline Street North, Waterloo, ON, N2L 2Y5, CAN}
\altaffiltext{7}{Department of Physics and Astronomy, Northern Arizona University, Flagstaff, AZ 86011, USA}

\begin{abstract}
    Future space telescopes will directly image extrasolar planets at visible
    wavelengths.  Time-resolved reflected light from an exoplanet encodes
    information about atmospheric and surface inhomogeneities.  Previous
    research has shown that the light curve of an exoplanet can be inverted to
    obtain a low-resolution map of the planet, as well as constraints on its
    spin orientation. Estimating the uncertainty on 2D albedo maps has so far
    remained elusive.  Here we present \texttt{exocartographer}, a flexible
    open-source Bayesian framework for solving the exo-cartography inverse
    problem. The map is parameterized with equal-area HEALPix
    pixels.  For a fiducial map resolution of 192 pixels, a four-parameter
    Gaussian process describing the spatial scale of albedo variations, and two
    unknown planetary spin parameters, \texttt{exocartographer} explores a
    198-dimensional parameter space. To test the code, we produce a light curve
    for a cloudless Earth in a face-on orbit with a 90$^\circ$ obliquity. We
    produce synthetic white light observations of the planet: 5 epochs of
    observations throughout the planet's orbit, each consisting of 24 hourly
    observations with a photometric uncertainty of 1\% (120 data).  We retrieve
    an albedo map and---for the first time---its uncertainties, along with spin
    constraints.  The albedo map is recognizably of Earth, with typical
    uncertainty of 30\%.  The retrieved characteristic length scale is 88$\pm 7
    ^\circ$, or 9800 km.  The obliquity is recovered with a $1-\sigma$
    uncertainty of $0.8^\circ$. Despite the uncertainty in the retrieved albedo
    map, we robustly identify a high albedo region (the Sahara desert) and a
    large low-albedo region (the Pacific Ocean).
\end{abstract}

\section{Introduction}
Next-generation space telescopes like HabEx and LUVOIR promise to directly image nearby Earth twins in reflected light, allowing astronomers to measure the reflectance spectra of these planets \citep{Des_Marais_2002}, as well as to monitor their time-varying brightness \citep{Ford_2001}. We focus on this second goal, but note that it may be impossible to properly interpret spectra without context from time-resolved observations, and vice versa.  


Using only time-resolved photometry, we would like to infer a planet's spin and its albedo map \cite[for a recent review of exoplanet mapping, see][]{Cowan_Fujii_2017}.  Rotation and obliquity place strong constraints on the late stages of a planet's formation \citep[e.g.,][]{Schlichting_2007}. Rotation also determines the frequency of diurnal radiative forcing and the amplitude of Coriolis forces, while obliquity determines the latitudinal distribution of insolation and the amplitude of seasons.  Surface geography, on the other hand, bears witness to the geophysical/geochemical processes operating on a planet \citep[e.g.,][]{Abbot_2012, Cowan_Abbot_2014, Fujii_2014}, and---in the case of liquid water---can directly establish the habitability of a planet \citep{Robinson_2017}. Lastly, the spatial and temporal distribution of clouds is determined by global climate, orography, and large-scale circulation \cite[e.g.,][]{Showman_2013}.    


Much progress has been made since \cite{Ford_2001} showed that time-resolved photometry of Earth encodes information complementary to time-averaged spectroscopy.  First of all, researchers demonstrated how to use the brightness variations of a planet to estimate its rotational period \citep{Palle_2008,Oakley_2009}.  It was then shown that the rotational \emph{color} variations of a planet can be used to infer the number, reflectance spectra, and longitudinal locations of major surface types \citep{Cowan_2009,Cowan_2011a, Fujii_2010, Fujii_2011, Cowan_Strait_2013,Fujii_2017}. Meanwhile, the rotational and orbital color variations of an unresolved planet can be analyzed to create a 2-dimensional multi-color map---equivalently a 2D map of known surfaces---and measure rotational obliquity \citep{Kawahara_2010,Kawahara_2011,Fujii_2012,Schwartz_2016,Kawahara_2016}.

\cite{Kawahara_2010} demonstrated the retrieval of a surface albedo map and the obliquity from simulated 1-year light curves of an atmosphere-less Earth. Mapping the surface from light curves is an ill-conditioned inverse problem that is unstable to noise. To overcome it, they adopted the physical condition that the surface albedo is between 0 and 1 and recovered the rough surface features of Earth. The obliquity was measured by minimizing $\chi^2$ or the extended information criterion, along with its uncertainty by bootstrap resampling. 

\cite{Kawahara_2011} and \cite{Fujii_2012} applied the concept of 2D mapping to simulated light curves of a realistic cloudy Earth, with an updated inversion technique. In these studies they employed Tikhonov regularization for the albedo map instead of the bounding condition. This method discards the components associated with small singular values of the design matrix and is equivalent to adopting a Gaussian prior in a Bayesian framework. Their recovered 2D surface maps at different photometric bands exhibit features consistent with the actual maps of cloud/snow cover, continents, and vegetation. However, evaluating the uncertainty of the recovered map and obliquity was left aside.

In this paper, we improve on the work of \cite{Kawahara_2011} in three important ways: 1) We base our map on only five epochs of 1 day each, rather than a year's worth of exposures, 2) we fit for the characteristic length scale of albedo markings on the planet, rather than adopting an a priori spatial scale, and 3) we quantify the uncertainty on the albedo map.

\section{\exocart{}}
\exocart{} is a fully Bayesian framework for retrieving the albedo map and spin geometry of a planet based on time-resolved photometry. The Python code is open-source and available at \url{https://github.com/bfarr/exocartographer}. 

\subsection{The Forward and Inverse Problems}
We cannot hope to extract the same detailed information \emph{from} the light curves as went \emph{into} them.  We will instead use a relatively simple surface integral that captures the essential physics governing the light curves.  Following \cite{Cowan_2013}, we describe the time-resolved reflectance of the planet as:
\begin{equation}\label{forward}
R(t) = \oint A(\theta, \phi) K(\theta, \phi, \mathbb{S}, t) \sin\theta d\theta d\phi,
\end{equation} 
where $A(\theta, \phi)$ is the 2D albedo map of the planet, which we assume to be fixed, $K(\theta, \phi, \mathbb{S}, t)$ is the convolution kernel, $\theta$ is co-latitude, $\phi$ is longitude, and $\mathbb{S}$ represents planetary spin parameters that are not known \emph{a priori}, namely obliquity and its orientation with respect to the observer. Computing $R(t)$ given everything on the right hand side of (\ref{forward}) is the forward problem. 

The inverse problem is to determine $A(\theta, \phi)$ and $\mathbb{S}$, given $R(t)$, the photometric uncertainty $\sigma$, and a parametrization of $K$.  In practice, this entails repeatedly solving the forward problem with varying parameters to see which ones best match the data in  hand. In order for this to be computationally feasible, one must make simplifying assumptions: the model used for retrieving the albedo map and planetary spin is essentially a toy model, albeit one that captures the first-order physics.     
We adopt the kernel for diffuse reflection from \cite{Cowan_2013}: $K = \frac{1}{\pi}V(\theta, \phi, t) I(\theta, \phi, t)$, where $V$ and $I$ denote the visibility and illumination functions. All of the time-dependence ---and the dependence on planetary spin--- enter the forward problem through the visibility and illumination.  They can be expressed compactly in terms of the angles between the local normal and the vector pointing towards the observer and the star: $V = \max(\cos\gamma_o, 0)$ and $I = \max(\cos\gamma_*, 0)$, where $\gamma_o$ and $\gamma_*$ are the observer and stellar zenith angles, both of which are a function of time and location on the planet.   

The visibility and illumination functions are more usefully expressed in terms of latitude and longitude of the sub-observer and sub-stellar positions: 
\begin{equation}\label{visibility}
  V =  \max \left\{\def\arraystretch{1.2}%
  \begin{array}{l}
    s\theta s\theta_o (c\phi c\phi_o + s\phi s\phi_o) + c\theta c\theta_o\\
    0 \\
  \end{array}\right.
\end{equation}
\begin{equation}\label{illumination}
  I =  \max \left\{\def\arraystretch{1.2}%
  \begin{array}{l}
    s\theta s\theta_s (c\phi c\phi_s + s\phi s\phi_s) + c\theta c\theta_s\\
    0 \\
  \end{array}\right.
\end{equation}
where we have used $s$ and $c$ to denote sine and cosine, and the $o$ and $s$ subscripts denote the sub-observer and sub-stellar location, which are functions of time. The computational crux of the forward problem is thus to quickly compute the sines and cosines of these time varying angles as a function of the orbital and spin parameters \citep[e.g., Appendix A of][]{Schwartz_2016}.




\subsection{Basis Maps}
In order to make the inverse problem tractable, we need to adopt a parametrization for the albedo map, $A(\theta, \phi)$.  As discussed in \cite{Cowan_Fujii_2017}, there are two complementary classes of basis maps one can adopt: pixels and spherical harmonics.  For the current application it is necessary to switch back and forth between these two representations to take advantage of both of their strengths.  

Spherical harmonics are an orthonormal basis, and complete for any continuous map on a sphere.  This means that the coefficients we derive for an expansion up to, say $l=3$ should remain unchanged if we extend the expansion to higher $l$. (In practice this is only strictly true if one is decomposing a map rather than a light curve, but it is roughly true for light curves, too: adding higher-order spherical harmonics should only produce small changes in the lower-order coefficients). Moreover, the spherical harmonic basis set exhibits a null-space: certain maps have no light curve signature.  By using spherical harmonic basis functions, we can trivially quantify the extent to which our Gaussian Process prior constrains otherwise unconstrained coefficients. The spherical harmonic representation is convenient for mapping because it enables coherent jumps in large regions of the planet, and enables a straightforward application of regularization, as described below.  

Pixels are also an orthonormal basis.  They have a more intuitive nullspace (e.g., latitudes more than $\pi/2$ away from the sub-observer latitude are simply unobservable and hence in the nullspace).  The pixel representation is convenient because the albedo of a pixel must be between 0 and 1, which makes it easy to propose parameter jumps.

\begin{figure}[htb]
\includegraphics[width=\linewidth]{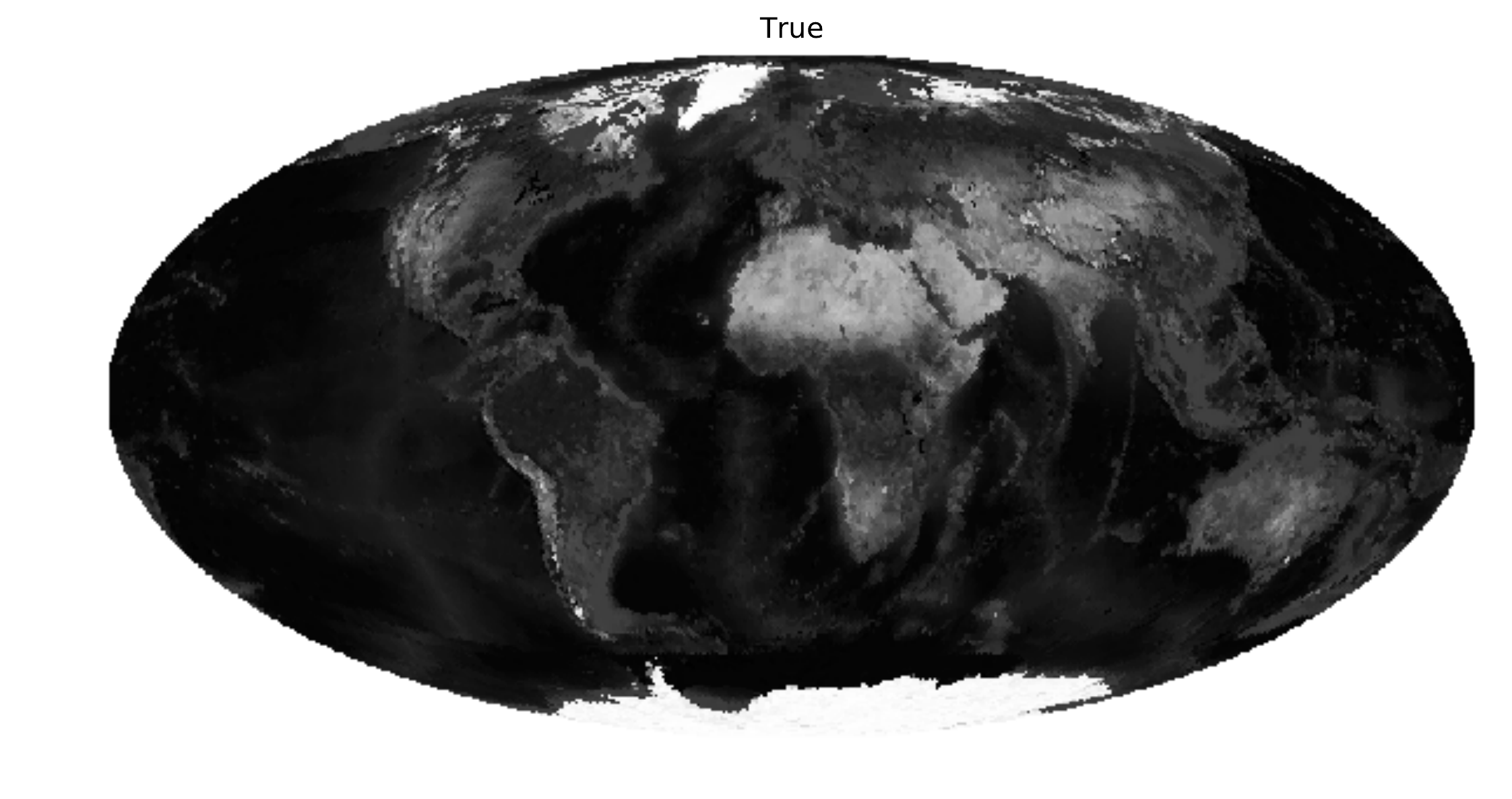}
          \includegraphics[width=\linewidth]{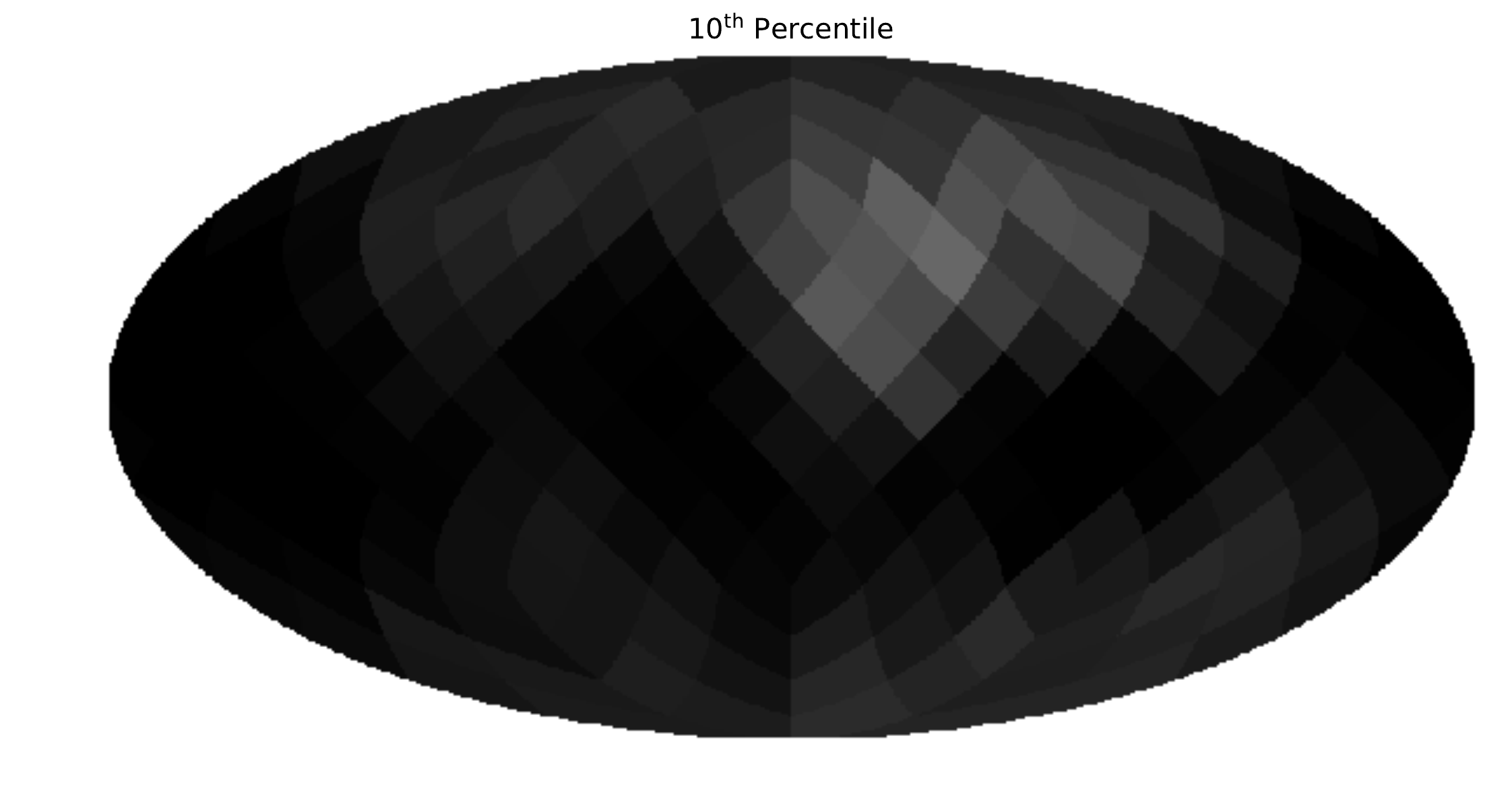}
          \includegraphics[width=\linewidth]{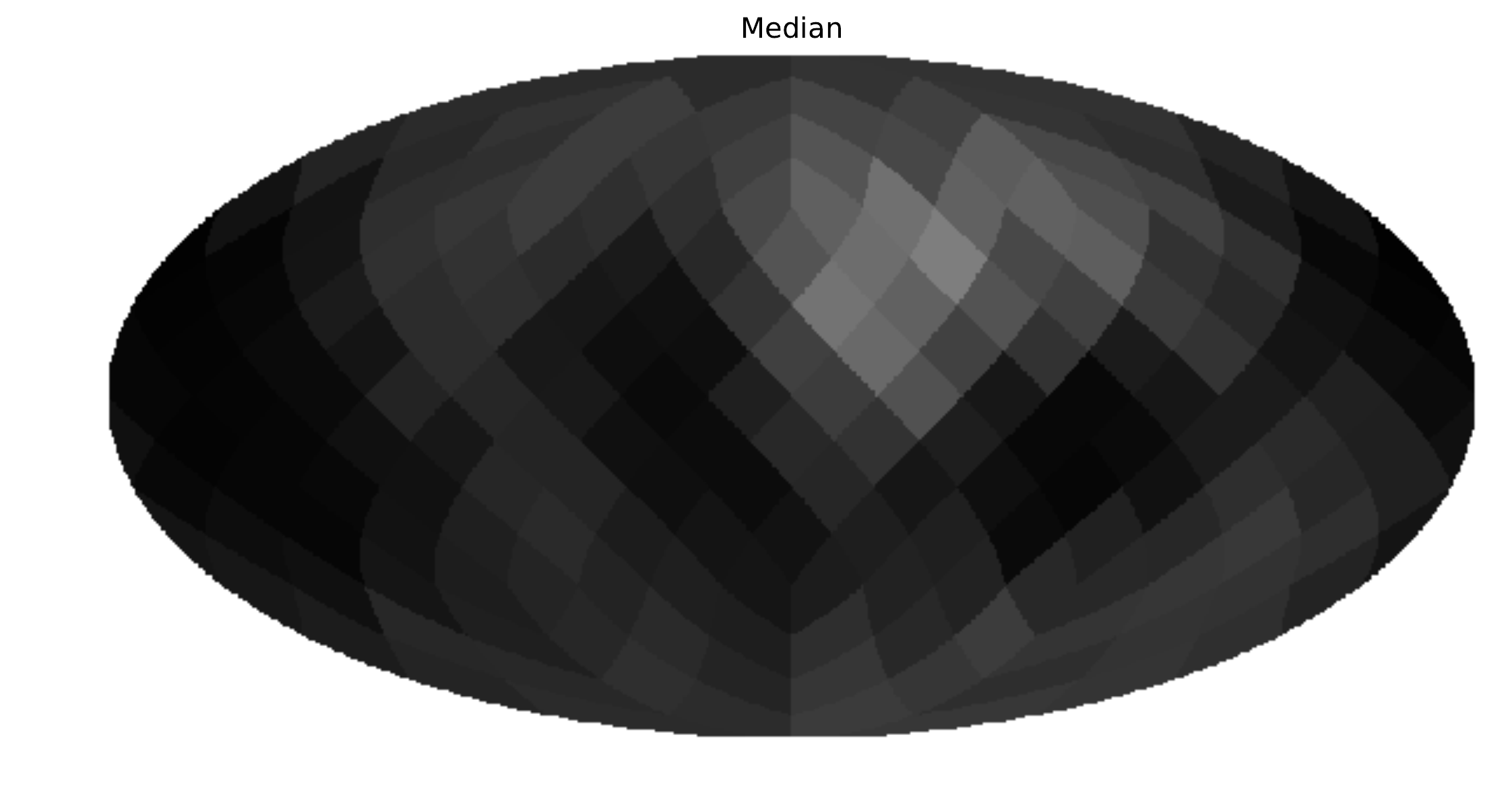}
          \includegraphics[width=\linewidth]{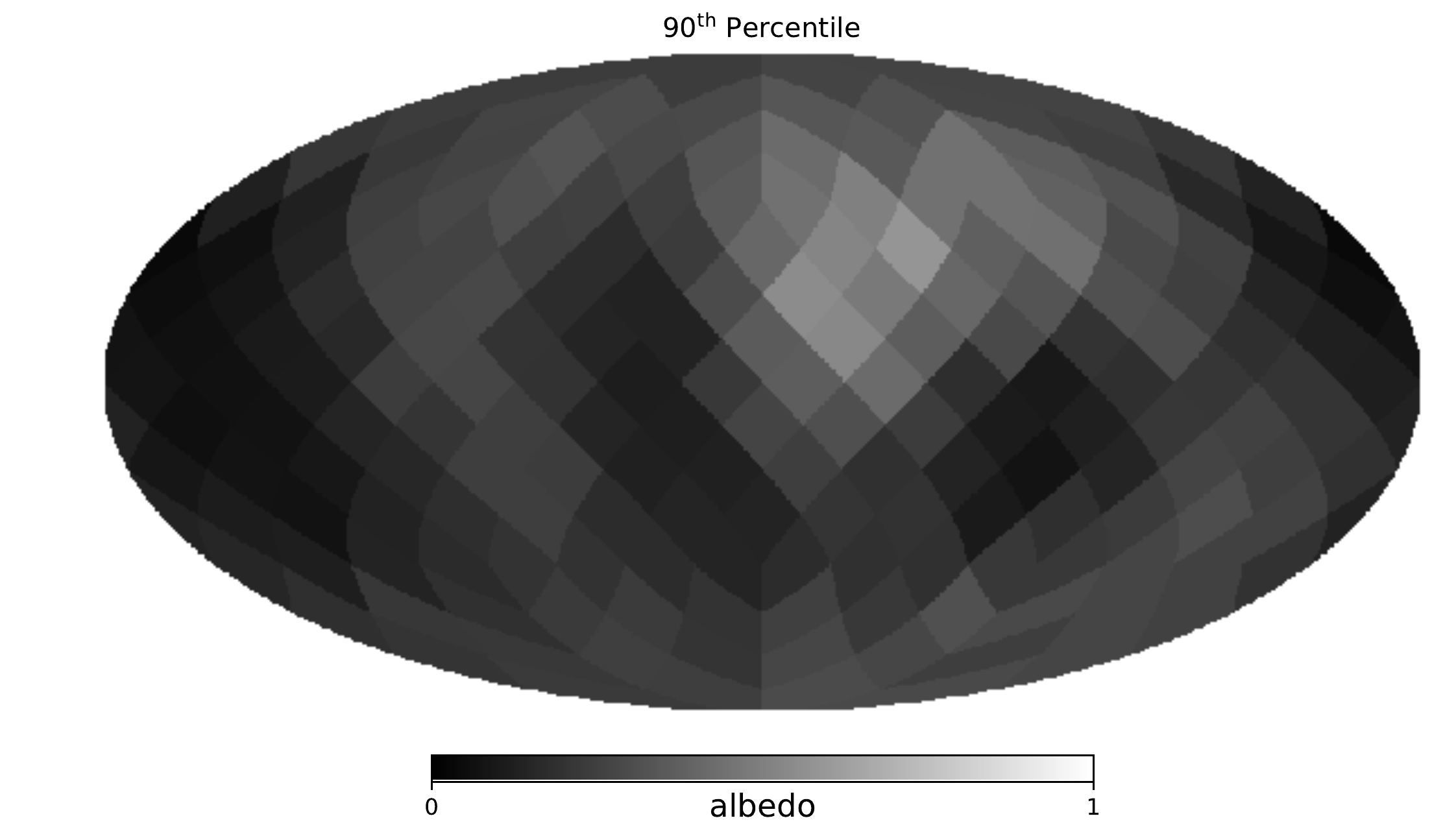}
\caption{The top map shows the true map used to generate the light curve, down-sampled to the resolution of our model.  This is followed by the $10^\mathrm{th}$ percentile, median, and $90^\mathrm{th}$ percentile of the marginal posterior distributions for the albedo of each pixel. The 10$^{\rm th}$ percentile map clearly shows a reflecting region (the Sahara Desert) while the 90$^{\rm th}$ percentile clearly shows a large dark region (the Pacific Ocean).  These maps therefore establish the presence of continents and oceans on the the planet, with all that implies for habitability. \label{fig:maps}}
\end{figure}

We use Hierarchical, Equal Area, and isoLatitude
Pixelation (HEALPix)\footnote{http://healpix.sourceforge.net/} pixels, so our map parameters are the albedo of each pixel (see Figure~\ref{fig:maps} for an example). HEALPix is superior to a regular lat-lon grid because all pixels are the same area, nothing is weird at the poles, and it plays well with spherical harmonics.  It takes a considerable amount of time to output the latitude and longitude of the HEALPix pixels, so we do this once ahead of time.  Since the latitude and longitude only appear in Eqns~\ref{visibility} and \ref{illumination} as trigonometric functions, we actually pre-compute $\{\cos\theta, \sin\theta, \cos\phi, \sin\phi\}$. 

The base HEALPix resolution is 12 pixels. Higher resolutions are defined by $N_{\rm side}$, the number of divisions along the side of a base pixel. The number of HEALpix pixels in a map is therefore $N_{\rm pix} = 12 N_{\rm side}^2$.   We test the necessary resolution of the kernel by producing light curves with the same map, but with kernels of different resolution.  We find that $N_{\rm side} = 8$ and 4 introduce pixelation error near quadrature of $10^{-3}$, and a few $\times10^{-3}$, respectively.  Photometry of directly imaged planets is unlikely to be better than 1\% in the foreseeable future \citep{Cowan_2009, Fujii_2012}, so we deem that  $N_{\rm side} = 4$ is sufficient for our purposes (Fujii \& Kawahara 2012 use $N_{\rm side} = 8$). 

One can devise a heuristic for the $N_{\rm side}$ one should adopt as a function of orbital phase.  The kernel is a lune and therefore has a height of $\pi$ and a width $\pi-\alpha$, where $\alpha$ is the star-planet-observer ``phase angle''.  Each pixel has an angular area of $4\pi/N_{\rm pix}$, and an angular size of $\sqrt{4\pi/N_{\rm pix}} = \sqrt{\frac{\pi}{3}} N_{\rm side}^{-1} \approx N_{\rm side}^{-1}$. In order to properly resolve the kernel, one would like at least three pixels across the narrow width of the kernel:  $N_{\rm side}^{-1} \lesssim (\pi-\alpha)/3$, suggesting that the minimum resolution at a given orbital phase is $N_{\rm side} \gtrsim 3/(\pi-\alpha)$. At quadrature ($\alpha=\pi/2$), for example, we find that the minimum $N_{\rm side}$ is $6/\pi \approx 2$, justifying our use of $N_{\rm side} = 4$.  

Our map parameters are the pixel albedos.  The kernel of the convolution is computed on the Healpix map, so the transformation from map to light curve is simply matrix multiplication. But the Gaussian Process prior is applied on the spherical harmonic coefficients, $a_l^m$, so we must convert from pixels to spherical harmonics at each step in the MCMC.  (At first blush, it seems that one could decompose the kernel into spherical harmonics as well and perform the convolution in $Y_l^m$ space, but this is no faster because there are roughly as many spherical harmonics as there are pixels.)  Unfortunately, one cannot have exactly the same number of pixels as spherical harmonics.  In particular, for $N_{\rm side} = \{1,2,4,8\}$ the number of pixels is $N_{\rm pix} = \{12, 48, 192, 768\}$.  The number of spherical harmonics is $N_{SH} = (l_{\rm max}+1)^2$, which corresponds to $N_{SH} = \{9, 16, 36, 49, 169, 196, 729, 784\}$ for $l_{\rm max} = \{2, 3, 5, 6, 12, 13, 26, 27\}$.  Our chosen Healpix resolution of $N_{\rm side} = 4$ roughly translates to $l_{\rm max} = 13$.  Since $N_{SH}$ is slightly greater than $N_{\rm pix}$, we are slightly over-constraining the map, as described below.

\subsection{Gaussian Process}
Our adopted resolution of $N_{\rm side}=4$ means that we have 192 map parameters, namely the albedo of each pixel.  This is more than the 120 data we will fit below, so we must apply additional constraints on the pixel values. The Gaussian Process (GP) prior on the map imparts a preferred length/angular scale to the albedo structures, which could correspond to the typical size of clouds or continents on a given planet.  Unlike standard regularization, we do not have to choose that angular scale a priori, rather it is a fitted parameter along with the pixel values themselves and the viewing geometry parameters. 

At each step in the fit, all of the pixels, viewing parameters, and GP parameters are varied independently.  We then use the current GP model to evaluate the prior on the current map.  The GP parameters themselves have priors, so there are two layers of priors, making this a ``hierarchical'' model.  The GP parameters and their priors are: the mean albedo (flat prior), the standard deviation of the albedo (prior is flat in log), the preferred angular scale (flat prior between $1/3$ of a pixel to $3\pi$), and the relative amplitude of spatially-uncorrelated albedo variations (flat prior between 0 and 1). Since the priors on the GP parameters only indirectly affect the posterior, these choices are not critical.

The posterior, the probability of parameters, $\theta$, given data, $d$, is given by:
\begin{align}
p(\theta|d) = & p(\theta_{\rm GP}) p(\theta_{\rm map}|\theta_{\rm GP}) \nonumber \\ 
& \times p(\theta_{\rm orb}) p(d|\{\theta_{\rm map}, \theta_{\rm orb}\}),
\end{align}
where the last term is the likelihood of the data given the map, orbital, and Gaussian process parameters.

Evaluating the GP prior (particularly the multivariate Gaussian likelihood) in pixel space is expensive, so we take advantage of the property that any Gaussian map whose correlation matrix depends only on angular separation between points will have a diagonal covariance matrix in $Y_l^m$ space (i.e., the spatial analog of stationarity in the time domain; stationary processes are completely described by their Fourier-space power spectrum, which is the diagonal of the covariance matrix in Fourier space). Furthermore, any map that is statistically isotropic will have a $Y_l^m$ covariance that is only a function of $l$, which we will refer to as $C_l$. So, instead of computing the covariance matrix and using a multivariate Gaussian likelihood in pixel space, we compute the $a_l^m$ coefficients, and use a Gaussian likelihood in Spherical Harmonic space.  Each $a_l^m$ is distributed like $a_l^m \sim N\left(0, \sqrt{C_l}\right)$.

There is only one final wrinkle: the number of degrees of freedom in $a_l^m$ coefficients is not equal to the number of pixels, as described above. We can either: 1) under-constrain the pixels by using an $l_\mathrm{max}$ that corresponds to fewer $a_l^m$ coefficients than pixels, which will result in some linear combinations of pixels (corresponding to the $Y_l^m$ with $l > l_\mathrm{max}$) that are unconstrained by the prior; or 2) over-constrain the pixels by choosing an $l_\mathrm{max}$ that corresponds to more $a_l^m$ coefficients than pixels, meaning the effective prior imposed in pixel space is not exactly the squared-exponential kernel. We choose to do the latter, so the pixels are over-constrained. This makes sampling easier, since no combinations of pixels can run away to infinity.


\subsection{Viewing Geometry}
We assume that all of the parameters specifying the viewing geometry are known a priori, except for the planet's spin orientation (obliquity, and it's orientation with respect to the observer).  The remaining parameters can either be constrained by other means (the orbital inclination and period, as well as the planet's instantaneous location in its orbit can be constrained by radial velocity, stellar astrometry, or planetary astrometry), or are arbitrary (we don't worry about offending extraterrestrials by redefining their Greenwich). We presume that the planetary rotation period is already known.

\subsection{Parameter Sampling}
For a given light curve, we use the maximum likelihood estimator Powell to determine the best-fit map and viewing geometry, especially the planet's spin.  We use this as the initial guess for a thorough exploration of parameter space to determine parameter uncertainties.  

In order to explore parameter space, \exocart{} uses the Markov Chain Monte Carlo \emph{emcee} \citep{Foreman-Mackey_2013}.  In particular, we use parallel tempering \citep{Vousden_2016} to improve sampling efficiency of the high-dimensional, non-linearly correlated posterior. 

\begin{figure*}[htb]
\includegraphics[width=\linewidth]{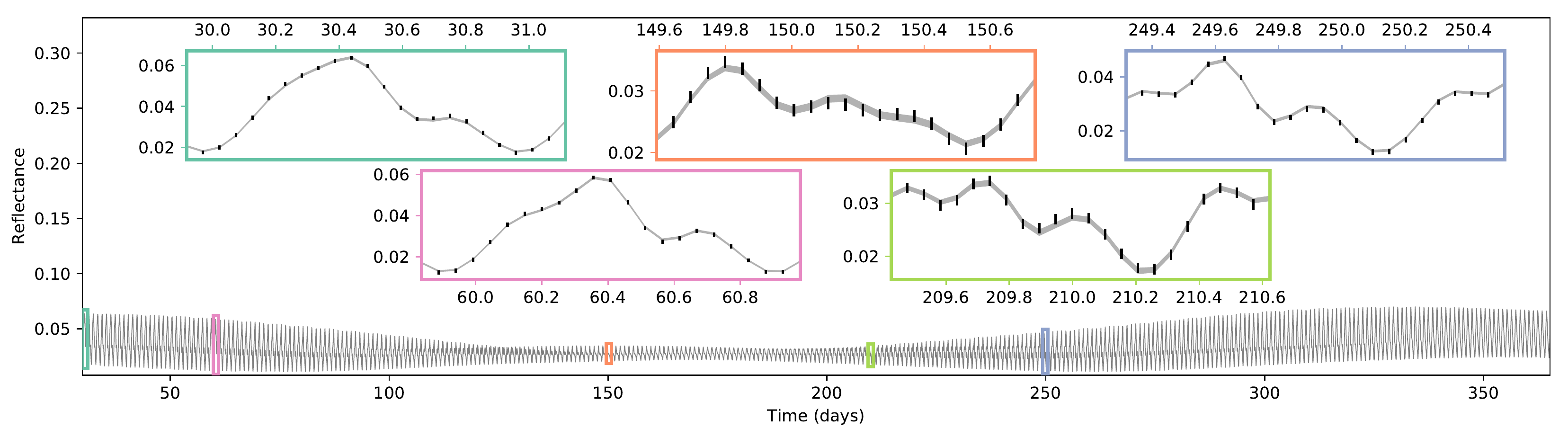}
\caption{Simulated light curve and resulting fit from an Earth analog with $0$ degree (face-on) inclination and $90$ degree obliquity.  The simulated data and measurement errors are shown in the five insets, representing five epochs, each lasting one rotation, collected over the course of a year with a one hour cadence.  The shaded region in each inset shows the central $90\%$ posterior credible interval for the reconstructed light curve. \label{fig:lightcurve}}
\end{figure*}

\section{Testing \exocart{}}
\subsection{Synthetic light curve}
We produce an idealized light curve with hourly cadence using a cloud-free toy model of Earth assuming Lambertian reflection (a higher resolution version of the top panel of Figure~\ref{fig:maps}) and adopting a face-on orbit ($i=0$) and 90 degree obliquity, but otherwise Earth-like values (1~day rotation, and 365~day orbit). The synthetic light curve is shown in Figure~\ref{fig:lightcurve}.  This is the most favorable viewing geometry for exoplanet mapping: 1) the planet--star projected separation is constant and hence the planet is visible throughout its orbit, 2) since it is always at quadrature, scattering phase effects can be neglected, 3) the planet is viewed equator-on and therefore the entire planet can be mapped, and 4) the large obliquity ensures that all latitudes are well illuminated at at least one point in the orbit, making it possible to recover a faithful map of the entire planet.  Indeed, this idealized scenario is precisely what Kawahara \& Fujii (2010) adopted for their seminal paper. 








Leveraging the work of \cite{Schwartz_2016} we only consider 5 epochs, and in deference to the fact that these missions will be horribly over-subscribed, we only observe the planet for slightly more than a planetary rotation at each epoch.  Considering 5 days---rather than a year's---worth of data has the ancillary benefit of reducing the run time of forward model calls by almost two orders of magnitude.  Generating a light curve across 5 days with a 1 hour cadence from a 192-pixel map takes 1.6 ms with a 3 GHz Intel Xeon W processor, and a full posterior probability density evaluation (including light curve generation) takes 2.3 ms.

\subsection{Retrieval Results}
\begin{figure}[htb]
\includegraphics[width=\linewidth]{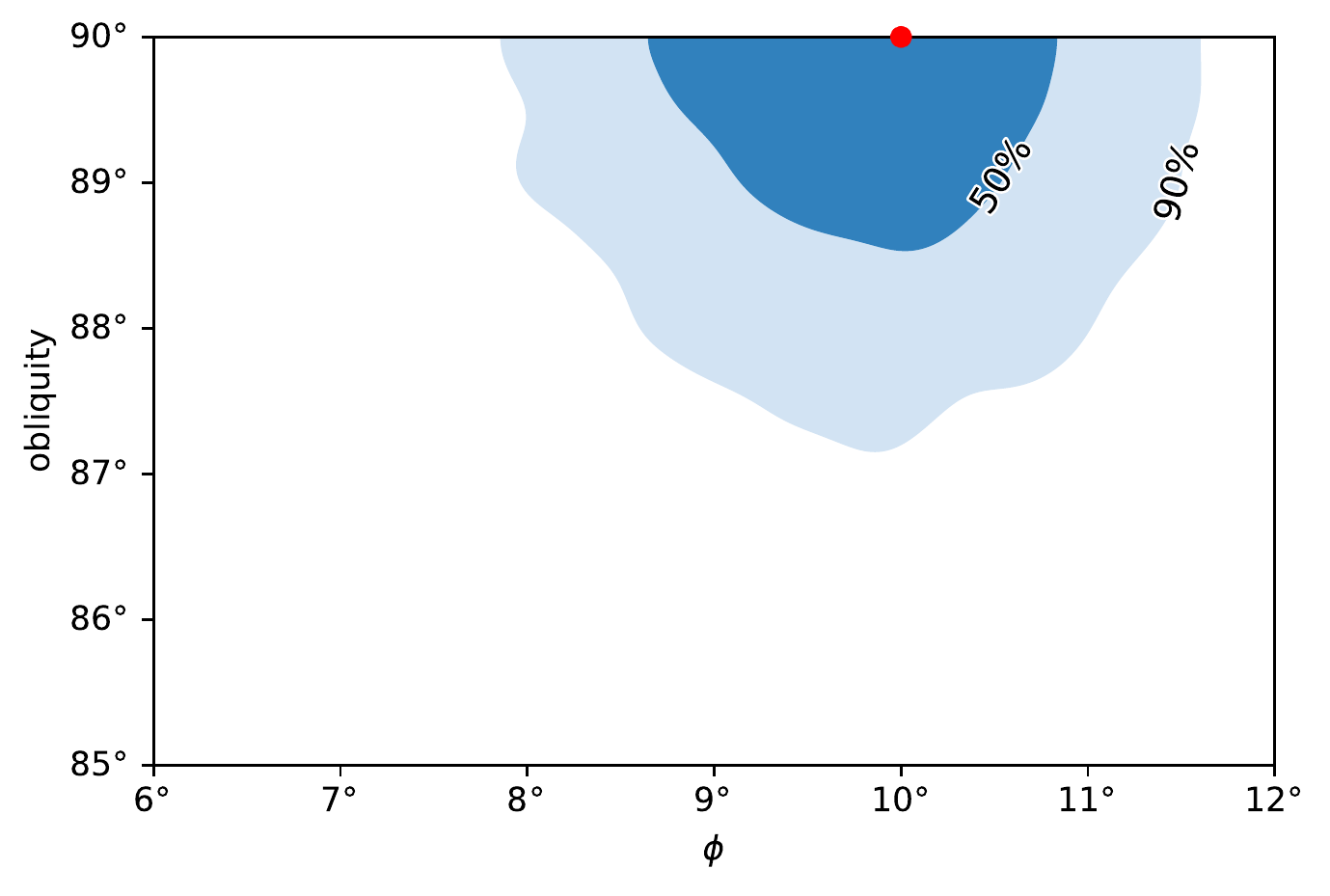}
\caption{Posterior samples (black) compared to true (red) values for the planet's spin orientation, parameterized here as obliquity and an azimuthal angle $\phi$. The retrieved spin orientation has an uncertainty of $\sim$2$^\circ$ and is consistent with the injected value of 90$^\circ$ and 10$^\circ$. \label{fig:spin}}
\end{figure}

The retrieval exercise was a surprising success.  We can successfully model the simulated photometry (Figure~\ref{fig:lightcurve}), but this is nothing to write home about: as stated above, the problem would be under-constrained if it weren't for the Gaussian Process prior.  More importantly, using only 5 days worth of photometry, we recover the obliquity and its orientation to high precision (Figure~\ref{fig:spin}), and the albedo map of the planet with enough precision to robustly identify a high-albedo region (the Sahara Desert) and a low-albedo region (the Pacific Ocean). The 10$^{\rm th}$ percentile, mean, and 90$^{\rm th}$ percentile maps are shown in Figure~\ref{fig:maps}. Although previous mapping efforts had proven that the best-fit map bears a resemblance to Earth \citep{Kawahara_2010,Kawahara_2011,Fujii_2012}, we have now shown that---even with very limited orbital coverage---the map uncertainties are small enough to make robust inferences about the surface character of the planet.

\section{Discussion \& Future Work}
We have built an open-source, modular code that builds on \cite{Fujii_2012}.  Crucially, we use an MCMC to extract albedo maps and planet spin \emph{and their uncertainties}, whereas they used optimization to determine best-fit parameters.  Also, we use Gaussian Processes to enforce smooth maps rather than Tikhonov Regularization.  The latter has a tunable regularization parameter, $\lambda$, which is chosen based on the ``L-curve criterion''.  By contrast, Gaussian Processes uses the data to fit for the characteristic length scale of the map.  

Although this was a useful numerical experiment, we made many simplifying assumptions, all of which should eventually be relaxed in future efforts. \exocart{} is a solid platform on which to build more sophisticated mapping capabilities. We now discuss possible improvements, starting with relatively straightforward considerations and ending with more challenging improvements.  

\subsection{Low-Hanging Fruit}
Our assumed photometric uncertainty of 1\% for one hour integrations is at the optimistic end of what LUVOIR might achieve for nearby targets ($\sim$10~pc). Nonetheless, smaller telescope diameters might be able to achieve comparable photometric precision for planets somewhat larger than Earth that rotate more slowly. In any case, larger photometric uncertainties will produce larger parameter and map uncertainties, but in no way stress the \exocart{} platform. 

We assumed a known rotational period for the planet, whereas in practice this would have to be extracted from the data themselves. Previous efforts have shown that this can be done independently of exo-cartography \citep[e.g., by computing the auto-correlation function of the time-resolved photometry;][]{Palle_2008,Oakley_2009}. It is unlikely that an MCMC initialized near the correct rotational period will jump to one of its harmonics, unless the time sampling is very poor, in which case planet mapping is hopeless.

By forcing the albedo to be between 0 and 1, we have implicitly assumed that the planetary radius is known.  In reality, the radii of directly imaged planets will not be known. Allowing for unknown planetary radius effectively removes the $A\le 1$ constraint: is the planet very reflective or simply very big?  Even in that case there is a physical limit to the size of planets, roughly that of Jupiter. The $A>0$ constraint would be unaffected by an unknown radius: flux cannot be negative, regardless of planet size. For the particular retrieval excercise we have performed, the albedo of cloudless Earth is quite low, so we expect that our results would be robust to within a scale factor; we wouldn't know the absolute albedo of the Sahara, merely that it is some factor greater than the planetary mean.  

The face-on geometry is favorable for mapping because the planet is never inside the coronagraph/starshade inner working angle, provided it is visible at all. But since we only considered five epochs for our analysis, considering different orbital inclinations should not present a significant challenge. That said, considering more, or longer, epochs would increase the number of data, hence increasing the computational burden but ultimately  improving the quality of the maps and geometrical constraints.

The 90$^\circ$ obliquity of the planet facilitated our task in two ways: all latitudes receive significant sunlight at one or more orbital phases, and all latitudes are visible to the observer.  The combination of these two factors mean that we can, in principle, map the entire planet.  For different viewing geometries we would only be able to map certain latitudes (e.g., an observer at 45$^\circ$ north cannot map locations below 45$^\circ$ south) and some latitudes would be poorly constrained \citep[for low obliquity planets, the poles are poorly illuminated and hence hard to map, even for a polar observer;][]{Cowan_2011}. 

\subsection{Harder Nuts to Crack}
We have assumed that the surface of the planet is a non-uniform Lambertian reflector.  In fact, real planets are non-Lambertian, notably due to forward scattering from molecules and aerosols, as well as due to specular reflection by surface liquids \citep{Robinson_2010,Robinson_2014}. In principle one could use a parametrized scattering phase function to fit for this behavior.  Of course, the parameters would be different for different surface types.  Although this would add a few model parameters, the main challenge would be that forward model calls might become more computationally expensive due to the non-trivial convolution kernel. 

Lastly, we have adopted a cloudless planet. To first order, clouds are just another bright surface, and one can differentiate between clouds and, say, continents with multi-wavelength data---even two-broadbands can do the trick (Cowan et al. 2009; Fujii \& Kawahara 2012). But clouds further complicate the mapping exercise in two ways: they mask underlying surfaces, and they change with time. One can make a map of surfaces and average cloud cover \citep{Fujii_2012}, and cloud variability will show up as residual structured noise.  In this scheme, regions that are essentially always shrouded in clouds, e.g., rain forests, are treated as if the clouds were indeed glued to the surface. 

Is there a better way to deal with clouds? The very feature that makes clouds pernicious---their time-variability---may also be their undoing: since most regions are cloud-free, at least occasionally, then it may be possible to construct a cloud-corrected map.  This would require multiple rotations at each epoch, the duration of the observations must be greater than the characteristic weather timescale on the planet, \citep[e.g., a few days for Earth;][]{peixoto1992physics}.    

A possible strategy for dealing with both non-Lambertian scattering and clouds is using multi-wavelength data. In this manuscript we limited ourselves to white-light photometry.  Simultaneously analyzing multi-band data has the advantage that the maps can look very different at different wavelengths, while the geometrical parameters must be common \citep[e.g.,][]{Fujii_2012}.  One can even use the color variations of the planet to identify the number and colors of surfaces, even if these are a priori unknown \citep{Cowan_Strait_2013, Fujii_2017}.







\acknowledgements
We acknowledge the International Space Science Institute in Bern, Switzerland for hosting the series of three workshops that led to this paper. We thank Suzanne Aigrain, Yuka Fujii, Vikki Meadows, Enric Pall\'e, Joel Schwartz, and Ed Turner for helping with the workshop proposal and for fruitful discussions at the workshops.

HMH thanks the IGC at Pennsylvania State University for warm hospitality, Bard College for extended support to visit the ISSI with students, and the Perimeter Institute for Theoretical Physics for generous sabbatical support. This work is  supported  by  Perimeter  Institute  for  Theoretical  Physics.   Research  at  Perimeter  Institute is supported by the Government of Canada through Industry Canada and by the Province of Ontario through the Ministry of Research and Innovation.


\end{document}